\documentclass[conference]{IEEEtran}
\IEEEoverridecommandlockouts
\usepackage{cite}
\usepackage{amsmath,amssymb,amsfonts}
\usepackage{algorithmic}
\usepackage{graphicx}
\usepackage{amsthm}
\usepackage{physics}
\usepackage{placeins}
\usepackage{array}
\theoremstyle{plain}
\newtheorem{definition}{Definition}
\usepackage{textcomp}
\usepackage{xcolor}
\usepackage{soul}

\usepackage[colorinlistoftodos]{todonotes}

\def\BibTeX{{\rm B\kern-.05em{\sc i\kern-.025em b}\kern-.08em
    T\kern-.1667em\lower.7ex\hbox{E}\kern-.125emX}}
\begin{document}

\title{Scaling Qubit Mapping and Routing With Position Graph Abstraction and Memoization\\
}

\author{
\IEEEauthorblockN{
Brent Russon\IEEEauthorrefmark{1},
Bao G. Bach\IEEEauthorrefmark{1}\IEEEauthorrefmark{2},
Ed Younis\IEEEauthorrefmark{3},
Ilya Safro\IEEEauthorrefmark{2}\IEEEauthorrefmark{4}
}
\IEEEauthorblockA{\IEEEauthorrefmark{1}\textit{Quantum Science and Engineering, University of Delaware}, Newark, DE, USA}
\IEEEauthorblockA{\IEEEauthorrefmark{2}\textit{Computer and Information Sciences, University of Delaware}, Newark, DE, USA}
\IEEEauthorblockA{\IEEEauthorrefmark{3}\textit{Computational Research Division, Lawrence Berkeley National Laboratory}, Berkeley, CA, USA}
\IEEEauthorblockA{\IEEEauthorrefmark{4}\textit{Physics and Astronomy, University of Delaware}, Newark, DE, USA}
\IEEEauthorblockA{\{bhrusson@udel.edu, baobach@udel.edu, edyounis@lbl.gov, isafro@udel.edu\}}
}

\maketitle

\begin{abstract}
Scalable qubit mapping and routing remain major bottlenecks in quantum compilation, especially for Trapped-Ion Quantum Charge-Coupled device (TI-QCCD) architectures, where qubit interactions require physically shuttling ions under strict movement, congestion, and trap-capacity constraints. We present a compilation framework built around the position graph abstraction, a unified representation of executable locations, movement paths, and routing constraints that enables heuristic mappers to operate directly on shuttling-based hardware. Using this abstraction, we accelerate the SWAP-based BidiREctional heuristic search (SABRE) by implementing relative move scoring, which caches repeated heuristic move evaluations that arise during search, and memoized congestion resolution, which speeds up the resolution of repeated congestion. This optimization removes redundant computation without changing routing/shuttling decisions, improving the scalability of SABRE-based methods on TI-QCCD systems. Our results show that combining an architecture-aware abstraction with memoized heuristic evaluation yields a practical and effective path toward scalable qubit mapping and routing across heterogeneous quantum architectures.\\
Reproducibility: the source code and data are available at [link will be available here upon acceptance]
\end{abstract}


\section{Introduction}
Qudit mapping and routing are central bottlenecks in quantum compilation because hardware constraints often prevent logical two-qudit interactions from being executed directly\cite{zhu2025quantum, cowtan2019qubit}. This challenge is especially important for superconducting architectures and shuttling systems such as trapped-ion quantum charge-coupled device (TI-QCCD) architectures. In superconducting systems, physical qudits can be moved using SWAP operations along a given configuration \cite{ning2019deterministic}. In contrast, TI-QCCD requires physically shuttling ions through segments and junctions while respecting movement legality, trap-capacity limits, and congestion constraints \cite{malinowski2023wire}. As a result, routing overhead can dominate execution time and strongly affect scalability. 

Existing practical mapping methods are largely built around abstractions such as Coupling Graphs, which are well matched to SWAP-based routing on superconducting devices. However, architecture with qudit shuttling\cite{pino2021demonstration, siegel2024towards, de2025high}, such as TI-QCCD, requires reasoning not only about which qudits should interact, but also about where ions can legally move, which intermediate positions are occupied, and whether a candidate path is blocked. These constraints make standard connectivity-only abstractions insufficient for expressing the shuttling-based routing. Our prior work addressed this challenge by introducing the \emph{Position Graph} abstraction \cite{bach2025efficient}, a hardware-aware representation that captures executable locations, movement paths, and routing constraints within a unified graph model. Built on top of this abstraction, the SHAW and SHAPER \cite{bach2025efficient} compilers showed that heuristic search methods inspired by superconducting-qudit mapping can be adapted to shuttling-based trapped-ion compilation. Although this abstraction is expressive enough to capture the complex TI-QCCD hardware constraints, the initial implementation \cite{bach2025efficient} suffers from scalability issues where it can take a day to shuttle a $128$-qudit circuit on a $180$-ion space device.

This paper takes the next step toward the practical deployment of the Position Graph abstraction, which substantially improves the scalability, enabling large-scale shuttling with both Superconducting and TI-QCCD architectures. We demonstrate that this richer abstraction does not incur meaningful overhead compared to previous methods, and highlight performance improvements to BQSkit's SABRE \cite{li2019tackling, younis2021berkeley} and SHAW algorithms by using cached results of architectural information modeled by the Position Graph. These include finding all-pairs shortest paths, their distance matrix, and an executable subgraph. While the Position Graph enables architecture-aware routing, heuristic search over QCCD states can still incur substantial overhead because the same local blockage and congestion situations are evaluated repeatedly during routing. To reduce this overhead, we introduce \emph{LightSHAW}, a cached variant of
SHAW's congestion-resolution procedure. LightSHAW reuses architecture-dependent
information such as movement paths and local clearing regions, and it avoids
recomputing local congestion scores when the same occupied local configuration
recurs during recursive blockage clearing. These caches reduce repeated graph
traversal and local occupancy scanning while preserving SHAW's routing
decisions. 

In addition to improving SHAW, we evaluate the runtime cost of using the Position Graph abstraction in a standard superconducting routing setting. A concern with richer architectural models is that they may impose unnecessary overhead when the target device is well described by a traditional Coupling Graph. To test this directly, we instantiate LightSABRE using both a conventional Coupling Graph backend and a Position Graph backend. The two implementations perform the same routing decisions and produce identical compiled circuits, allowing the comparison to isolate the cost of the abstraction itself. Our results show that the Position Graph can match the runtime of the Coupling Graph while retaining the ability to represent richer hardware structure when needed. This suggests the practicality of the Position Graph, maintaining runtime while potentially serving as a unification for different types of architectures (we already showed for SWAP-based and shuttling-based architectures).

In summary, this paper makes the following contributions:
\begin{enumerate}
    \item We identify and remove redundant computations in SHAW by
    caching shortest MOVE paths, travel-time
    distances, and blockage-related path information used repeatedly
    during routing.

    \item We introduce LightSHAW, a cached variant of SHAW that memoizes
    local congestion geometry and occupancy-sensitive congestion scores,
    reducing repeated work during recursive blockage resolution while
    preserving the routing behavior of the original algorithm.

    \item We demonstrate that the Position Graph abstraction can replace
    the traditional Coupling Graph for LightSABRE-style routing with
    negligible runtime overhead, showing that a richer architecture model
    need not come at the cost of slower compilation.

    \item We provide an open-source implementation of the Position Graph
    framework and its routing machinery, supporting experimentation across
    both superconducting and TI-QCCD compilation workflows and future extension
    across heterogeneous quantum architectures.
\end{enumerate}


\begin{figure}
    \centering
    \includegraphics[width=\linewidth]{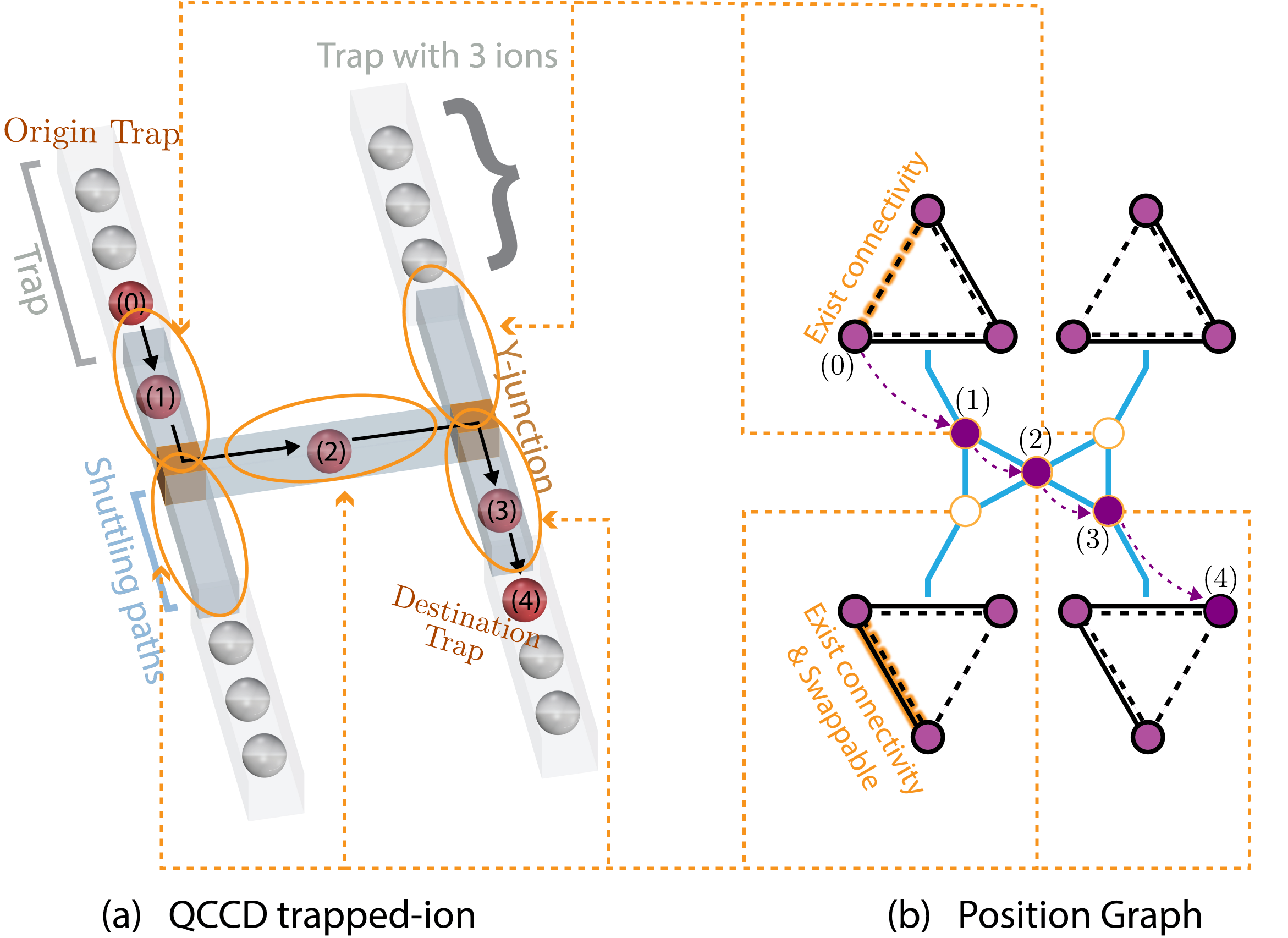}
    \caption{Example of mapping a QCCD-based TI architecture to its corresponding Position Graph.
    Figure (a) shows a real hardware configuration, and Figure (b) shows the Position Graph abstraction that represents the hardware configuration. Here, (i) each \textit{position} vertex in the Position Graph represents a physical location where an ion may reside. The vertex colored purple indicates that an ion is currently occupying this location, while the uncolored (white) vertex indicates that this position is empty. (ii) Each vertex with a black border represents the available position inside a trap, while the vertex with a yellow border represents abstract intermediate positions where a qudit may temporarily reside while moving between traps. (iii) The dotted black edges within each trap indicate pairs of positions between which a two-qudit gate operation is possible, while the straight line indicates the physical connection that allows operations such as swap. (iv) The blue edges abstract the path where the ion can move (without stopping) to reach another trap.}
    \label{fig:main_fig}
\end{figure}


\section{Background}
\label{sec:QIC}
\subsection{Quantum Compilation}
In quantum computing, compilation is the process of converting a logical quantum circuit into an equivalent circuit that can be executed directly on a specific quantum processing unit (QPU). This requires adapting the circuit to hardware-dependent constraints, including restricted qudit connectivity, imperfect gate fidelities, and other architectural limitations. Effective compilation is especially important in the NISQ setting, where noise and short coherence times make resource efficiency critical. As a result, quantum compilation generally aims to reduce both circuit size and execution time in order to improve overall reliability. The two central stages of this process are typically (1) transpilation and (2) qudit mapping and routing.

\paragraph{Transpilation}
The purpose of transpilation is to transform a quantum circuit into an equivalent representation expressed entirely in the native gate set of a target QPU. Because different hardware platforms support different universal gate libraries, this step is necessary to make a circuit physically executable on a given device. For example, superconducting architectures \cite{koch2007charge} commonly use gates such as $SX$, $RZ$, and $CNOT$, whereas trapped-ion platforms \cite{kielpinski2002architecture} may instead employ $RZ$, $U1q$, and $RZZ$.  State-of-the-art methods for transpilation \cite{younis2022quantum} often operate by partitioning a large circuit into smaller subcircuits, computing the unitary implemented by each block, and then re-synthesizing each block in the target native gate set using an appropriate synthesis procedure

\paragraph{Qudit Mapping and  Routing}
Qudit mapping, also referred to as layout, is the compilation stage that assigns the logical qudits of a circuit to the physical qudits of a target architecture while respecting hardware constraints. This problem is most commonly discussed for superconducting platforms, where connectivity is described by a Coupling Graph and multi-qudit gates can only be executed between connected physical qudits. Similar placement requirements also arise in trapped-ion and neutral-atom systems, where each logical qudit must be assigned to a specific ion or atom \cite{zhu2025quantum}. Once an initial layout is chosen, qudit routing \cite{bapat2023advantages} is used to realize interactions between distant logical qudits by inserting additional operations—typically SWAPs, shuttling operations, or other movement primitives—that bring the relevant qudits into an executable configuration. In superconducting architectures, these routing transitions are typically implemented through SWAP operations between adjacent physical qudits. QCCD-based trapped-ion systems face an analogous routing problem, but the mechanism differs: instead of exchanging logical states through SWAP gates, the compiler coordinates ion shuttling operations that physically move qudits across the device. Mapping and routing are inherently coupled: the quality of the initial assignment strongly affects the routing overhead, while routing costs in turn influence which assignments are desirable. Because these additional operations can significantly increase circuit depth, gate count, and execution time, mapping and routing are major sources of compilation overhead. Accordingly, modern methods \cite{li2019tackling, bhattacharjee2019muqut, liu2023tackling, zou2024lightsabre, yan2024quantum} aim both to find high-quality initial placements and to perform intermediate remapping steps that enable future gates at low cost.

\section{Related Work}

\subsection{Heuristic Mapping and Routing}
Heuristic methods form the dominant practical approach to qudit mapping and routing, primarily because they scale more effectively than exact or solver-based formulations.
For superconducting devices, SABRE~\cite{li2019tackling} established the standard heuristic template based on front-layer routing with lookahead. It maintains a front layer of executable gates and an extended lookahead layer, and iteratively inserts SWAPs according to a heuristic that balances immediate and future routing costs; the same procedure can also be applied bidirectionally to obtain improved initial layouts. Recently, PAM~\cite{liu2023tackling} extended this framework by combining block-based routing with permutation-aware synthesis, while LightSABRE~\cite{zou2024lightsabre} improved robustness, scalability, and routing quality on larger circuits.

In trapped-ion and especially QCCD architectures, however, heuristic methods must reason not only about logical remapping but also about physical ion transport. Early work therefore emphasized shuttle-aware initial placement and multi-trap compilation, including shuttle-efficient mapping policies and shuttle-reduction heuristics for trapped-ion systems \cite{murali2020architecting, saki2022muzzle}. Recent work has further developed architecture-specific scheduling heuristics for QCCD, including parallel shuttle strategies\cite{zhu2025s, ruan2025trapsimd}, or adapting the established SABRE method to QCCD architecture through abstraction \cite{bach2025efficient}.

\subsection{Analytical Mapping and Routing}
In addition to heuristic methods, a substantial line of work has studied qudit mapping and routing through explicit analytical formulations. In superconducting systems, graph-theoretic approaches have modeled the problem using subgraph isomorphism, token swapping, and routing on restricted-connectivity architectures, thereby clarifying its underlying combinatorial structure \cite{siraichi2019qubit, cowtan2019qubit}. Building on this view, several works formulate superconducting mapping and routing as exact optimization problems, including branch-and-bound methods for exact qudit allocation, SMT-based layout synthesis, and integer linear programming models \cite{zhu2020exact, tan2020optimal, nannicini2022optimal}.

Similar analytical directions have recently appeared for trapped-ion architectures. For example, \cite{tseng2024satisfiability} proposes an SMT-based formulation for qudit mapping in one-dimensional trapped-ion systems, whereas \cite{schoenberger2024using} formulates exact shuttling in QCCD architectures (considering grid memory zone and single processing zone) as a Boolean satisfiability problem to minimize the required number of movement timesteps. \cite{bach2025efficient} derived a mixed-integer linear program that fully captures the dynamics of QCCD architecture as described in \cite{murali2020architecting} through their Position Graph abstraction. These approaches are valuable because they yield rigorous optimization baselines and, in some settings, provably optimal schedules, although their runtime usually restricts them to relatively small or moderately sized problem instances. \cite{zhang2021time}.

\section{GROUNDWORK}

\subsection{Position Graph Abstraction}

We adopt the Position Graph abstraction introduced in~\cite{bach2025efficient} as presented in Figure \ref{fig:main_fig}.
In that work, the abstraction is defined concretely for trapped-ion QCCD architectures. 
We restate this formulation below and then describe how the abstraction extends to 
other hardware models.

\begin{definition}[Position Graph]
Given a hardware architecture consisting of (1) a set of ion traps $T$, 
with $|T| = m$, where each trap $t \in T$ has capacity $n_t$ ions, 
(2) a set of junctions $J$, and (3) a set of segments $S$, we define 
the Position Graph $G_p = (V, E, \psi)$, where $V$ is the set of vertices, 
$E$ is the set of edges, and $\psi : E \rightarrow L_E$ is a labeling 
function that assigns labels to edges from the set $L_E$.

A node $i \in V$ corresponds to a possible position of an ion in the 
device at a given time. An edge $(i,j) \in E$ corresponds to a possible 
transition of an ion from position $i \in V$ to $j \in V$.

The number of vertices and edges are given by:
\begin{equation}
|V| = \sum_{t \in T} n_t + |S|
\end{equation}
\begin{equation}
\begin{aligned}
|E| &= \sum_{t \in T} (n_t - 1) 
+ \sum_{j \in J} \frac{d(j)\big(d(j) - 1\big)}{2} \\
&\quad + |J_T| - |J_J|,
\end{aligned}
\end{equation}
where $j$ denotes a junction, $J_T$ is the set of segments connecting 
junctions and traps, $J_J$ is the set of segments connecting pairs of 
junctions, and $d(j)$ denotes the degree of junction $j$.

The set of edge labels is defined (for QCCD architectures) as:
\begin{equation}
L_E = \{\texttt{swap},\ \texttt{merge/split},\ \texttt{move}\}.
\end{equation}

The \texttt{swap} label corresponds to transitions within a trap, 
\texttt{merge/split} corresponds to transitions between trap and transport 
regions, and \texttt{move} corresponds to transitions within transport regions.
\end{definition}

While Definition~1 provides a concrete construction for QCCD systems, 
the Position Graph abstraction itself is more general. At its core, it is a 
labeled graph $G_p=(V,E,\psi)$ in which vertices represent occupiable 
physical locations and edges represent valid transitions or interactions 
between locations. The specific sets $T$, $J$, and $S$ define one particular 
instantiation of this abstraction rather than a requirement of it.

This interpretation enables a unified representation across different hardware 
architectures. For superconducting systems, the Position Graph simplifies 
considerably: each vertex corresponds directly to a physical qudit and is executable, and edges 
represent allowed two-qudit interactions. In this case, there are no intermediate 
transport regions (i.e., no junctions or segments), and transitions correspond 
only to gate execution. Under this specialization, the Position Graph reduces 
to the well-known Coupling Graph, while preserving the ability to model richer 
architectural features when needed.

To represent the dynamic state during routing or shuttling, the Position Graph is paired 
with a placement function
\[
\Phi : Q \rightarrow V,
\]
which maps each logical qudit in the set $Q$ to a physical position in $V$. 
This separation between the static hardware structure ($G_p$) and the dynamic 
assignment ($\Phi$) enables efficient reasoning about routing, scheduling, 
and resource constraints.

\subsection{SHAW: A Shuttling-Aware Heuristic for QCCD Architectures}

SHAW (SHuttling-AWare heuristic)~\cite{bach2025efficient} is a
SABRE-based routing and mapping algorithm designed for trapped-ion QCCD architectures.
Whereas SABRE routes circuits by inserting SWAP operations on a fixed coupling
graph, SHAW routes by selecting physically valid ion-shuttling operations on a
Position Graph.

In superconducting architectures, a gate is executable when its qudits are
adjacent according to the Coupling Graph. In QCCD architectures, a gate is
usually executable when the participating ions are co-located within an executable trap.
SHAW therefore replaces the SABRE objective of bringing qudits adjacent with
the QCCD objective of bringing ions into the same trap.

At a high level, SHAW follows the same iterative structure as SABRE. The
algorithm maintains a front layer of gates that are ready to execute. If a
front-layer gate is immediately executable, SHAW executes it and advances the
frontier. Otherwise, it evaluates candidate shuttling moves and selects the move
that minimizes a heuristic cost over the front layer and a small lookahead set.

The cost model reflects the physical constraints of QCCD hardware. Rather than
only minimizing graph distance, SHAW evaluates moves according to whether they
bring the required ions closer to the same executable trap, reduce the distance
among the ions participating in a gate, and respect trap capacity constraints.
The Position Graph provides the static hardware structure, while the current ion
assignment records, which physical positions are occupied during shuttling.

Thus, SHAW demonstrates that SABRE-based heuristic routing can be extended from
static qudit adjacency constraints to dynamic ion placement and trap co-location constraints. The next
subsection describes how SHAW handles cases where movement and occupancy
constraints create congestion.

\subsection{Congestion Resolution in SHAW}

Congestion arises in SHAW when an ion must move along a shuttling path, but an
intermediate position on that path is already occupied. SHAW refers to such an
occupied intermediate position as a blockage. When a movement path is blocked,
SHAW invokes a congestion-resolution procedure to clear the blockage and allow
shuttling to continue.

When congestion resolution is required for a front-layer gate, SHAW first
selects an executable target trap for that gate. Candidate traps are evaluated
using the estimated cost of moving the gate ions into the trap, together with
penalties for congested paths when positions along the required shuttling paths are occupied. Traps with
more available space and less congested paths are preferred because they are less likely to introduce
additional congestion.

After selecting the target trap, SHAW assigns the participating ions to
positions in that trap and attempts to move each ion along a shortest shuttling
path. If the next position on the path is unoccupied, the ion moves directly. If
the next position is occupied, SHAW attempts to clear the blocking ion.

To clear a blockage, SHAW evaluates nearby positions (except positions on the dedicated path) as possible destinations for the blocking ion. These candidate positions are scored using a local
congestion metric: positions whose nearby regions contain fewer occupied
positions are preferred. The size of the nearby region is controlled by a search
depth. A depth of one considers positions one step away from the
candidate position, a depth of two includes positions up to two steps
away, and so on. In the implementation used in \cite{bach2025efficient}, the initial search
depth is one less than the trap capacity, and the depth increases during
recursive congestion-resolution calls.

If an available nearby position exists, the blocking ion is moved there; this is called a clearing move. If all nearby positions are themselves occupied, SHAW recursively applies the same
procedure to clear one of those secondary blockages. In difficult cases, the
resolver may temporarily move the target ion itself to escape a local dead end.

Congestion resolution is therefore a recursive local repair mechanism. It does
not change SHAW's overall SABRE-based search structure; rather, it allows the
search to continue when QCCD path, occupancy, and trap-capacity constraints
prevent a direct shuttling move.

\section{Methodology}

\subsection{Quantum Architecture Caching}

Most qudit routing and mapping algorithms across different architectures can use
the Position Graph as the central architecture representation. Since the
architecture does not change during a compilation run, architecture-dependent
quantities can be computed once and reused throughout routing. These quantities include an all-pairs distance matrix for the positions in the Position Graph and the corresponding shortest paths list. Using the Position Graph's edge labels and selecting the edges annotated with execute functionality, we can derive and cache an execution subgraph from the original Position Graph. This execution subgraph is used for efficiently determining whether a multi-qudit gate can be performed. 

In our implementation, this design exposed repeated computations in both BQSKit's SABRE implementation \cite{younis2021berkeley} and the original SHAW implementation \cite{bach2025efficient}, including repeated distance and path calculations, and repeated checks on allowed gate execution. Moving these architecture-level queries into Position Graph caches reduced this overhead without changing the routing
decisions. The remainder of this section focuses on the additional caching introduced for the implementation of the SHAW algorithm, which we will refer to as LightSHAW from here on. LightSHAW also introduces a SHAW-specific global cache of path blockage profiles and local caches that help make SHAW's congestion resolution more efficient. These caches are specific to the TI-QCCD architecture.

\subsection{LightSHAW Global Path Blockage Profiles}
In our LightSHAW implementation, we introduce a new global cache of path blockage profiles for each source-target movement path. A path blockage profile is the list of intermediate positions along a selected movement path that could block the move if occupied, together with the fixed penalty associated with resolving a blockage at each such position.

For a movement from position $s$ to position $t$, the profile depends only on
the Position Graph and on the fixed costs assigned to physical shuttling
operations. It does not depend on the current ion assignment. During scoring,
LightSHAW checks the current assignment only to determine which positions in the profile are occupied.

This cache is useful because SHAW repeatedly scores candidate moves between the
same pairs of positions. By caching the path and the fixed blockage penalties,
LightSHAW avoids reconstructing the same path and recomputing the same static
penalty terms on each scoring call. The cached profile does not determine that
a path is blocked; it only provides the reusable list of positions that must be
checked against the current ion assignment.

\subsection{LightSHAW Congestion Memoization}

Architecture-level caches are insufficient for congestion resolution because
congestion depends on the current ion assignment. When SHAW attempts to clear a
blocked movement path, it evaluates candidate clearing positions among the blocking ions by examining which nearby positions are currently occupied. These local congestion
evaluations can recur during recursive clearing.

LightSHAW preserves the congestion-resolution logic of SHAW but avoids
recomputing repeated local evaluations. It implements this optimization as a
two-level cache. The first level caches the local scoring set: for a given
candidate clearing position, it stores the positions that should be inspected
when computing congestion. The second level memoizes the resulting congestion
score for the current occupancy of that local scoring set.

\paragraph{Local Scoring-Set Cache}
During recursive congestion resolution, SHAW repeatedly evaluates nearby
locations as possible destinations for a blocking ion. For each candidate clearing position
$p$, target position $t$, blockage position $b$, and search depth $d$,
LightSHAW builds the local scoring set:
\[
    \mathcal{N}_d(p,t,b).
\]
This set contains positions reachable from $p$ within $d$ movement steps,
excluding the blockage position $b$ and the local target position $t$ for the
current clearing move. These exclusions keep the scoring region focused on
places where the blocking ion could be moved without immediately undoing the clearing move.
This set is then cached and does not decide which candidate position should be used. Instead,
for each candidate position already being considered by SHAW, it answers the
question: which positions should be inspected when scoring this candidate move?
LightSHAW caches this set under the tuple $(p,t,b,d)$ and reuses it whenever the same local structure is evaluated again.

\paragraph{Configuration-Keyed Congestion Memoization}
After retrieving the local scoring set: $\mathcal{N}_d(p,t,b)$ either from cache or by constructing the first time, LightSHAW checks which positions in that set are currently occupied. It does not store the full ion assignment in
the cache key. Instead, it forms a \emph{local occupancy signature} $\sigma$,
which records only the occupied positions among the candidate position $p$ and
the positions in $\mathcal{N}_d(p,t,b)$.

For example, suppose SHAW is evaluating candidate position $p=3$, and the
cached local scoring set is
\[
    \mathcal{N}_d(3,t,b)=\{4,5,7,9\}.
\]
If positions $5$ and $9$ are occupied, while positions $3$, $4$, and $7$ are
empty, then
\[
    \sigma=(5,9).
\]
If position $3$ were also occupied, then
\[
    \sigma=(3,5,9).
\]
Ions outside the candidate position and its local scoring set are not included because they do not affect the congestion score for this candidate move.

The congestion score is memoized using the key
\[
    (p,t,b,d,\sigma).
\]
On a cache miss, LightSHAW computes the same congestion quantities used by SHAW: the fraction of occupied positions in the local scoring set and a weighted
score that gives greater importance to occupied positions closer to the
candidate location. On a cache hit, the stored result is reused.

Because the key includes $\sigma$, the cached value remains valid even if ions
outside the local scoring set move. If the occupancy of the local scoring set
changes, then $\sigma$ changes, and the score is recomputed. The cache is scoped
to a single congestion-resolution episode and is passed through the recursive
calls generated while clearing a blocked movement path. This captures repeated local states during recursive clearing without allowing the cache to grow across the entire compilation. Together, these two caches preserve SHAW's congestion decisions while reducing
the repeated graph traversal and local occupancy scanning needed to score
candidate clearing moves.

\subsection{Pruning Trap Selection} 

During brute-force or congestion resolution, when a multi-qudit operation cannot be executed and gets stuck at the current ion assignment, the mapper must choose a candidate trap where the participating ions should be moved. An exhaustive implementation scores every executable trap exactly, but this is expensive because the exact score considers distinct assignments of ions to trap slots and includes congestion-resolution penalties along the selected paths, especially when the number of considered traps scales. We reduce this cost by first computing lower-bound scores for all traps. These traps are then evaluated as candidates in lower-bound score order, and exact scoring stops once no remaining lower bound can improve on the best exact score found so far.

In more detail, for each trap, we compute an optimistic lower bound on the cost of routing the participating ions into that trap. For each position in the position graph and each trap, we cache the shortest travel distance from that position to the closest position within the trap. The found score is the sum of these nearest-trap distances over the participating ions'
current positions. This bound ignores congestion and the requirement that different ions occupy distinct trap positions.

Each trap's score is then adjusted by its current occupancy: traps with more available positions receive a small bonus. Because the same adjustment is applied to both the lower bound and the exact score, the adjusted lower bound can only underestimate or equal the corresponding adjusted exact score.

Candidate traps are sorted by this lower bound. We then evaluate traps in that order using the exact scoring routine. Exact scoring considers the same shortest-path distances, but also adds congestion-resolution penalties for occupied positions along the paths. The exact scorer also enforces that each participating ion is assigned to a distinct position within the candidate trap.

As exact scores are computed, we maintain the best exact score found so far. Once the next candidate trap's lower bound is greater than the current best exact score, that trap cannot improve the solution. Since the candidate list is sorted by increasing lower bound, all remaining traps cannot improve the solution and are pruned without exact scoring.

\subsection{Discussion on memory usage for memoization}
The caching strategies described above trade memory for reduced repeated
computation. Let $\abs{V}$ denote the number of positions in the Position Graph,
$\abs{T}$ denote the number of executable traps, and $\mathrm{diam}(G_{p})$ denote the diameter of the position graph

The all-pairs distance matrix requires $O(\abs{V}^2)$ memory. The shortest path list has the worst-case memory as $ O(\abs{V}^2 \mathrm{diam}(G_{p}))$, since each source-target pair may store a path whose length grows with the architecture diameter.

The path blockage profile cache stores, for queried source-target movement
pairs, the intermediate positions along the path, and the fixed blockage
resolution cost for each such position. In the worst case, if every pair of
positions are queried, this cache also has the memory of $ O(\abs{V}^2 \mathrm{diam}(G_{p}))$. In practice, the actually memorys can grow much slower as it only considers the number of shuttling paths actually scored during compilation.

The trap-pruning cache stores the nearest distance from each position in the Position Graph to each executable trap. Its memory requirement is $O(\abs{V}\abs{T}).$

For a $k$-qudit operation, the lower-bound score for a trap is computed on
demand by summing $k$ entries from this cache. We intentionally avoid caching
the combined lower-bound score for tuples of participating ion positions,
because such a cache would scale as $O(\abs{V}^k \abs{T})$, which becomes prohibitive for large architectures.

Congestion memoization has two memory components. The geometry-only local
scoring-set cache stores neighborhoods of the form $\mathcal{N}_d(p,t,b)$, and scales with the number of unique local congestion geometries encountered:
\[
    O(NL),
\]
where $N$ is the number of cached $(p,t,b,d)$ tuples and $L$ is the average size
of the corresponding local scoring set. The occupancy-dependent congestion
score cache is scoped to a single congestion-resolution episode. Its memory
scales with the number of distinct local occupancy signatures encountered
during that episode and is discarded afterward.

These caches reduce repeated graph traversal, path reconstruction, and local
occupancy scoring. Their drawback is that the geometry-dependent caches can
grow substantially as the architecture size increases, especially caches with
quadratic dependence on $\abs{V}$. For this reason, LightSHAW caches reusable
single-position and single-path quantities, but avoids caching combined
multi-ion trap scores whose key space grows combinatorially.

\section{Experiments}
\label{sec:evaluation}
\subsection{Experimental setup}
Our experiments use a diverse benchmark suite consisting of well-known quantum algorithms and Hamiltonian simulation circuits with sizes ranging from 16 to 512 qudits. The suite includes the Quantum Approximate Optimization Algorithm (QAOA)~\cite{farhi2014quantum}, the Quantum Fourier Transform (QFT)~\cite{nielsen2001quantum}, and Hamiltonian simulation circuits for the Transverse Field Ising Model (TFIM) and Transverse Field XY (TFXY) model. QAOA is a variational algorithm for combinatorial optimization and is widely used as a representative NISQ workload. QFT is a core algorithmic primitive used in larger quantum algorithms, including Quantum Phase Estimation~\cite{abrams1999quantum} and Shor's factoring algorithm~\cite{shor1999polynomial}. TFIM and TFXY represent structured Hamiltonian simulation workloads, which are central to many near-term quantum computing applications.

The QFT and QAOA circuits are generated using Qiskit~\cite{qiskit2024}. For QAOA, we construct instances from Erd\H{o}s--R\'enyi graphs with edge probability $p=0.1$. The TFIM and TFXY circuits are generated using the F3C++ compiler~\cite{kokcu2022algebraic}.

To isolate the effect of the proposed Position Graph abstraction, we first compare LightSABRE~\cite{zou2024lightsabre} using the conventional coupling-graph representation against LightSABRE using our Position Graph abstraction. We then evaluate SHAW~\cite{bach2025efficient} and LightSHAW, as described in the methodology, to study the impact of architecture caching and congestion memoization in the shuttling-based setting. Together, these experiments serve as an ablation study that separates the contributions of the Position Graph abstraction from those of the LightSHAW memoization strategy.

We implement the Position Graph abstraction in Python~3.11 as an extension to BQSKit~\cite{younis2021berkeley}. We evaluate the scalability of LightSHAW against SHAW and QCCDSim~\cite{murali2020architecting}, a state-of-the-art QCCD shuttling simulator. All experiments are performed on a Gen3 Intel Xeon Gold 6240R system with 1~TB of main memory. Since QCCDSim only supports circuits transpiled to a gate set containing CNOT gates, we first transpile all input circuits to the required gate set using BQSKit to enable a fair comparison.\\
\noindent \emph{Reproducibility: Our source code and results are available at} [link will be added here upon acceptance].

\subsection{Experimental results}

\begin{figure*}
    \centering
    \includegraphics[width=\linewidth]{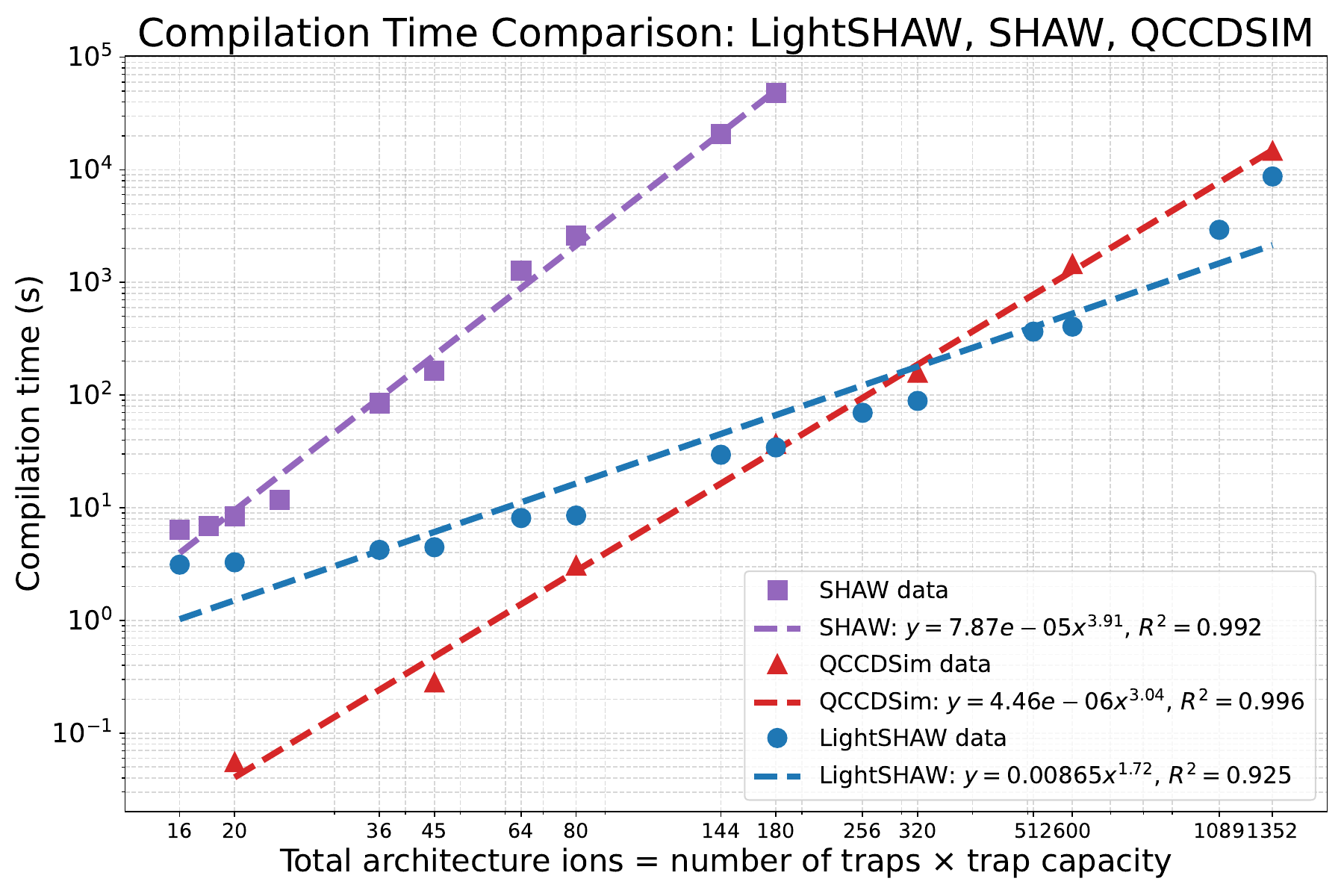}
    \caption{Average compilation time (s) of SHAW, LightSHAW, and QCCDSim across multiple benchmark circuits at different circuit sizes and architectures. The figure is shown on a logarithmic scale. The data point shows average compile times at a given architecture (Grid$x\times y$ with $k$ ions per trap). The dashed curves show power-law fits of the form \(y = a x^b\), where \(x\) is the total number of ions inside the architectures (total number of traps $\times$ Trap capacity). Both SHAW and LightSHAW can solve stricter architectures than QCCDSim, as shown in Table \ref{tab:can-exe}, which results in more data points than QCCDSim. The fitted exponents indicate the observed growth rates: SHAW scales approximately as \(x^{3.91}\), LightSHAW as \(x^{1.72}\), and QCCDSim as \(x^{3.04}\).}
    \label{fig:lightshaw_shaw_qccdsim_runtime}
\end{figure*}

\paragraph{QCCD Compilation time comparison}
We first evaluate the impact of the proposed caching techniques by comparing LightSHAW, SHAW, and QCCDSim compilation times through Figure~\ref{fig:lightshaw_shaw_qccdsim_runtime}. Our results show that LightSHAW significantly reduces compilation time from SHAW by eliminating redundant recomputation during shuttling. The LightSHAW compilation time results are more comparable to QCCDSim. LightSHAW presently takes additional time for shorter circuits, but when considering circuits on $180$-ion architecture, we start to match and outperform QCCDSim compilation runtime. Furthermore, the results show better compilation time scaling compared to SHAW when performing power law fitting model $y = a{x}^{b}$, where $x$ is the total number of qubits. 
For small to medium-sized benchmark (less than $100-$qubit circuits), QCCDSim consistently achieves lower compilation time, whereas LightSHAW takes longer compilation time due to additional overhead from its generalized abstraction and congestion-aware heuristic search. Despite this overhead, LightSHAW always has better performance, as shown in Table \ref{tab:QCCD-5ions}, and still shows a compilation time scaling advantage when comparing with QCCDSim, where we can see that the exponent factor $b$ for LightSHAW is around $1.72$, which is much better than QCCDSim, around $3.04$.

Besides considering the compilation time, Operation time (gate execution time and shuttling time) is a critical factor for near-term Trapped-Ion quantum devices. Following the gate fidelity model described in \cite{murali2020architecting}, longer operation times increase gate noise, which directly degrades circuit fidelity. Prior work has shown that reducing operation time is essential for improving reliability on noisy hardware~\cite{murali2020architecting, malinowski2023wire}.
We evaluate the quality of the resulting LightSHAW and QCCDSim compilations in terms of operation time in Table~\ref{tab:QCCD-5ions}.

Table~\ref{tab:QCCD-5ions} shows total operation time across different benchmarks with different architectures. As shown, LightSHAW consistently produces a schedule with much lower operation time compared to QCCDSim. This suggests that the caching strategy not only allows LightSHAW to have a better compilation time scaling than QCCDSim but also maintains the operation time improvement from the original SHAW. 

\begin{table}
\centering
\small
\setlength{\tabcolsep}{4pt}
\caption{Total operation time comparison for LightSHAW and QCCDSim on
grid architectures. Lower values correspond to shorter
execution time $+$ shuttling time and improved circuit efficiency.}
\resizebox{\linewidth}{!}{
\begin{tabular}{|l|c|c|r|r|r|}
\hline
Benchmark & Arch & \scriptsize Ions/Trap &  \scriptsize LightSHAW ($\mu$s) &  \scriptsize QCCDSim ($\mu$s) & Ratio \\
\hline
QAOA\_16 & $2\times 2$ & 5 & 21091 & 39680 & 0.532 \\
QFT\_16 & $2\times 2$ & 5 & 25327 & 53880 & 0.470 \\
TFIM\_16 & $2\times 2$ & 5 & 15603 & 39260 & 0.397 \\
TFXY\_16 & $2\times 2$ & 5 & 19171 & 29980 & 0.639 \\
\hline 
QAOA\_32 & $3\times 3$ & 5 & 45385 & 77100 & 0.589 \\
QFT\_32 & $3\times 3$ & 5 & 123523 & 193360 & 0.639 \\
TFIM\_32 & $3\times 3$ & 5 & 150577 & 182300 & 0.826 \\
TFXY\_32 & $3\times 3$ & 5 & 148751 & 160920 & 0.924 \\
\hline
QAOA\_64 & $4\times 4$ & 5 & 102357 & 158260 & 0.647 \\
QFT\_64 & $4\times 4$ & 5 & 323839 & 629820 & 0.514 \\
TFIM\_64 & $4\times 4$ & 5 & 692951 & 1193240 & 0.581 \\
TFXY\_64 & $4\times 4$ & 5 & 680465 & 1094320 & 0.622 \\
\hline
QAOA\_128 & $6\times 6$ & 5 & 432382 & 908440 & 0.476 \\
QFT\_128 & $6\times 6$ & 5 & 616460 & 1493960 & 0.413 \\
TFIM\_128 & $6\times 6$ & 5 & 3271462 & 3762880 & 0.869 \\
TFXY\_128 & $6\times 6$ & 5 & 3271462 & 3762880 & 0.869 \\
\hline
QAOA\_256 & $8\times 8$ & 5 & 1872099 & 5450600 & 0.343 \\
QFT\_256 & $8\times 8$ & 5 & 1736270 & 5171360 & 0.336 \\
\hline
QAOA\_512 & $10\times 10$ & 6 & 4705827 & 27545320 & 0.171 \\
QFT\_512 & $10\times 10$ & 6 & 1960315 & 9972900 & 0.197 \\
\hline
QAOA\_1024 & $13\times 13$ & 8 & 3525259 & 148271360 & 0.024 \\
QFT\_1024 & $13\times 13$ & 8 & 4492742 & 21024420 & 0.213 \\
\hline
\end{tabular}
}
\label{tab:QCCD-5ions}
\end{table}
\FloatBarrier

Beyond improved performance, similar to SHAW, LightSHAW demonstrates significantly improved robustness. As shown in Table~\ref{tab:QCCD-4ions}, QCCDSim fails to produce valid
compilations for all the strict architectures where the congestion is high (denoted by X). In contrast, LightSHAW consistently produces valid compilations for these examples and can do so across a broader range of circuits and architectures. This indicates that LightSHAW can significantly improve the practical executability of quantum circuits on noisy hardware.

\begin{table}
\centering
\small
\setlength{\tabcolsep}{4pt}
\caption{Operation time results for architectures with a stricter number of ions per trap (higher ion utilization). QCCDSim fails to produce valid compilations (denoted by X), while LightSHAW successfully completes all instances, demonstrating improved
robustness under tighter resource constraints.}
\resizebox{\linewidth}{!}{
\begin{tabular}{|l|c|c|r|r|}
\hline
Benchmark & Arch & \footnotesize Ions/Trap &   \footnotesize LightSHAW ($\mu$s) & \footnotesize QCCDSim ($\mu$s) \\
\hline
QAOA\_16 & $2\times 2$ & 4 & 29166  & X \\
QFT\_16 & $2\times 2$ & 4 & 34345 & X \\
TFIM\_16 & $2\times 2$ & 4 & 27727 & X \\
TFXY\_16 & $2\times 2$ & 4 & 28814 & X \\
\hline
QAOA\_32 & $3\times 3$ & 4 & 50118  & X \\
QFT\_32  & $3\times 3$ & 4 & 143611 & X \\
TFIM\_32 & $3\times 3$ & 4 & 173341 & X \\
TFXY\_32 & $3\times 3$ & 4 & 167387 & X \\
\hline
QAOA\_64 & $4\times 4$ & 4 & 105128 & X \\
QFT\_64  & $4\times 4$ & 4 & 335942 & X \\
TFIM\_64 & $4\times 4$ & 4 & 790249 & X \\
TFXY\_64 & $4\times 4$ & 4 & 777260 & X \\
\hline
QAOA\_128 & $6\times 6$ & 4 & 444328  & X \\
QFT\_128  & $6\times 6$ & 4 & 616460  & X \\
TFIM\_128 & $6\times 6$ & 4 & 3271462 & X \\
TFXY\_128 & $6\times 6$ & 4 & 3391871 & X \\
\hline
QAOA\_256 & $8\times 8$ & 4 & 1150680  & X \\
QFT\_256  & $8\times 8$ & 4 & 1285258  & X \\
\hline
QAOA\_512 & $8\times 8$ & 8 & 4647942  & X \\
QFT\_512  & $8\times 8$ & 8 & 1639341  & X \\
\hline
QAOA\_1024 & $11\times 11$ & 9 & 18817281  & X \\
QFT\_1024  & $11\times 11$ & 9 & 3408264  & X \\
\hline
\end{tabular}
}
\label{tab:QCCD-4ions}
\end{table}

Overall, these results show that QCCDSim remains faster for small to medium instances where it successfully solves. On the other hand, LightSHAW provides a compilation approach capable of handling a wider range of circuits while maintaining practical compilation time, and even better for big instances, producing schedules with much better operation time.


\paragraph{Compatibility of Coupling Graph and Position Graph shown on LightSABRE}

To evaluate the overhead of the Position Graph abstraction in a setting where a traditional Coupling Graph is sufficient, we implemented LightSABRE using both abstractions. The Coupling Graph backend serves as the standard
superconducting-device baseline, representing the architecture through physical
adjacency and distance. The Position Graph backend uses a similar logic,
but stores architecture information in a richer labeled graph.

At the time of testing, BQSKit's SABRE algorithm did not have the relative scoring described by LightSABRE implemented. To enable better scaling and experiments with larger circuits, we implemented the relative scoring technique with both the Coupling Graph and Position Graph implementations. We were able to implement the Position Graph such that the same SWAP operations are chosen as the version featuring the Coupling Graph. This allows for a more direct measurement of the overhead of the Position Graph. 

Profiling revealed two architecture-level queries that were repeatedly computed
in the Coupling Graph implementation. First, the original executability check
validated multi-qudit gates by constructing an induced subgraph and checking
whether it was connected. Although a single subgraph construction is inexpensive,
this check is called millions of times on larger circuits. Table~\ref{tab:can-exe}
shows the impact of replacing repeated subgraph construction with the same information cached by the Position Graph.

\begin{table*}
\centering
\small
\setlength{\tabcolsep}{4pt}
\caption{Profiling of \texttt{\_can\_exe} implementations in LightSABRE.
CG evaluates executability using repeated subgraph extraction, whereas PG
uses cached Position Graph execute adjacency. Values are averaged over two
runs.}
\begin{tabular}{|l|l|r|r|r|r|}
\hline
Benchmark & Method & Total (s) & \texttt{\_can\_exe} (s) &
\# Calls & \texttt{get\_subgraph} (s) \\
\hline
QAOA\_128 & CG & 41.69 & 14.06 & 254445 & 11.38 \\
QAOA\_128 & PG & 28.84 & 2.13 & 254445 & -- \\
\hline
QFT\_128 & CG & 21.69 & 5.75 & 93206 & 4.69 \\
QFT\_128 & PG & 14.06 & 0.84 & 93206 & -- \\
\hline
QAOA\_256 & CG & 318.57 & 111.27 & 1914174 & 89.98 \\
QAOA\_256 & PG & 220.89 & 16.90 & 1914174 & -- \\
\hline
QFT\_256 & CG & 95.26 & 24.82 & 403538 & 20.17 \\
QFT\_256 & PG & 51.92 & 3.49 & 403538 & -- \\
\hline
QAOA\_512 & CG & 2375.96 & 816.92 & 11786673 & 659.05 \\
QAOA\_512 & PG & 1508.63 & 115.23 & 11786673 & -- \\
\hline
QFT\_512 & CG & 447.85 & 80.09 & 1149807 & 64.83 \\
QFT\_512 & PG & 193.94 & 11.51 & 1149807 & -- \\
\hline
\end{tabular}
\label{tab:can-exe}
\end{table*}

Second, the Coupling Graph implementation recomputed all-pairs shortest paths at
the beginning of each forward and backward SABRE pass. With two layout passes and the routing pass this results in five all-pairs computations. On the 512-qudit benchmarks, each
call takes roughly 40 seconds. In contrast, the Position Graph stores the
corresponding path and distance information with the architecture object and
reuses it across passes. Table~\ref{tab:all-pairs-profile} shows this profiling
difference.

\begin{table}
\centering
\small
\setlength{\tabcolsep}{4pt}
\caption{All-pairs shortest-path profiling comparison between Coupling Graph
(CG) and Position Graph (PG) representations in 512-qudit LightSABRE runs.
The Position Graph performs a single precomputation, whereas the coupling
graph recomputes shortest paths multiple times. Results are averaged over
two runs with 2 layout passes.}
\begin{tabular}{|l|l|>{\centering\arraybackslash}p{0.75cm}|r|r|}
\hline
Benchmark & Method & Calls & Cum. Time (s) & Total (s) \\
\hline
QAOA\_512 & CG & 5 & 210.62 & 2375.96 \\
QAOA\_512 & PG & 1 & 26.49 & 1508.63 \\
\hline
QFT\_512  & CG & 5 & 202.03 & 447.85 \\
QFT\_512  & PG & 1 & 27.17 & 193.94 \\
\hline
\end{tabular}

\label{tab:all-pairs-profile}
\end{table}

\begin{figure*}
    \centering
    \includegraphics[width=.95\textwidth]{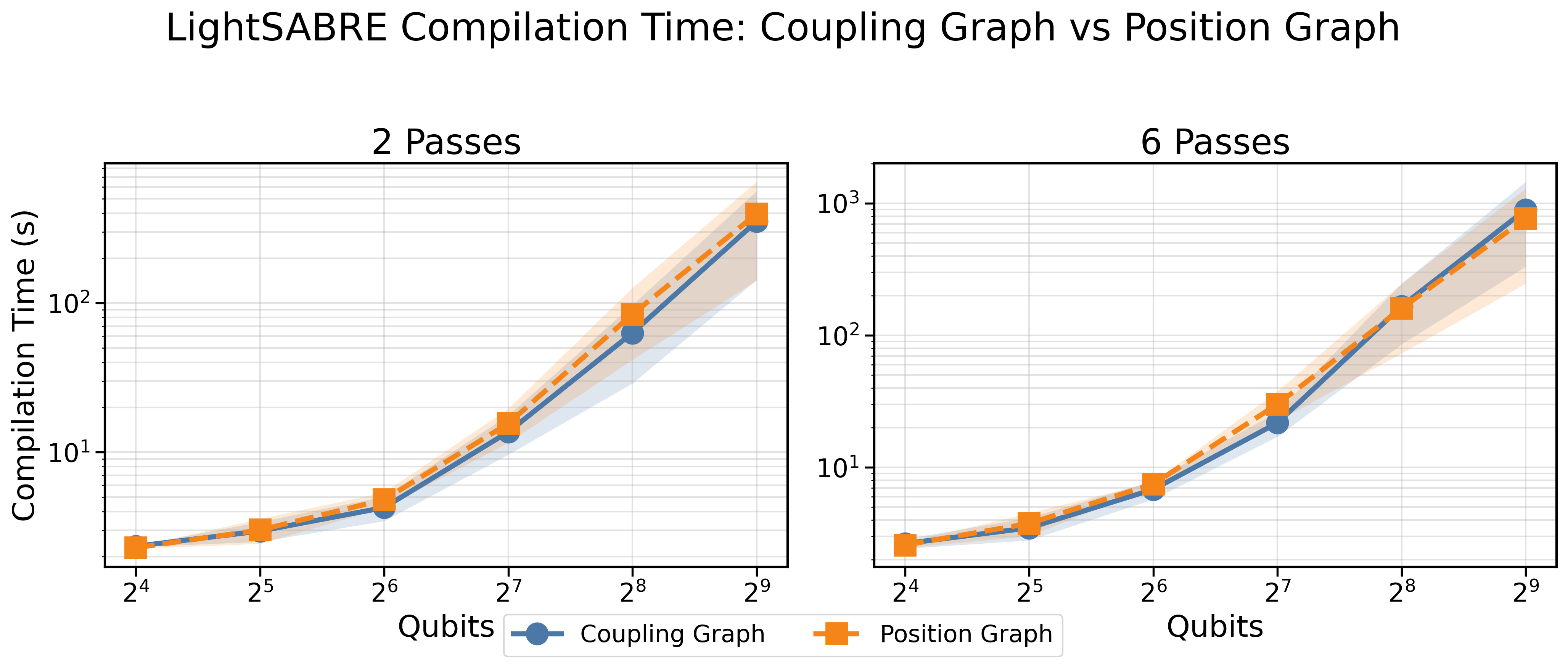}
    \caption{LightSABRE compilation time comparing Coupling Graph and Position
    Graph representations across QFT and QAOA benchmarks with varying
    qudit counts and 2 or 6 layout passes. This data includes caching strategies added to the Coupling Graph implementation. Results are shown on a logarithmic scale.
    Both representations produce identical compiled circuits, allowing a
    direct comparison of compilation time.}
    \label{fig:lightsabre-compilation-time}
\end{figure*}

These optimizations are not inherently exclusive to the Position Graph; similar
caches can also be added to a Coupling Graph implementation. The significance of
the result is that the Position Graph exposes these architecture-level queries
as part of the graph representation. As shown in Figure~\ref{fig:lightsabre-compilation-time},
when both back ends use comparable caching and relative scoring, the position
graph matches the compiliation time of the Coupling Graph with negligible overhead while preserving the ability to represent more types of architectures. This supports the Position
Graph as a practical common representation for both SWAP-based superconducting
routing and shuttling-based QCCD routing.

These results show that the Position Graph can replace the Coupling Graph for
SABRE-based routing without introducing meaningful overhead. When the target
architecture is simple, the Position Graph behaves like a Coupling Graph.
When the architecture is richer, the same object can expose movement paths,
execution regions, edge capabilities, and weighted costs. This allows a single
architecture representation to support both standard SABRE/LightSABRE for superconducting systems and SHAW/LightSHAW for more complex TI-QCCD systems.

\section{DISCUSSION}

A central implication of these results is that richer architecture
representations need not impose a prohibitive runtime cost. In the
superconducting-style LightSABRE experiments, the Position Graph matched the
behavior and runtime of a coupling-graph implementation while retaining
additional structure needed for QCCD routing. This suggests that compiler
infrastructure can move beyond minimal adjacency models without necessarily
sacrificing scalability.

Many routing workflows represent hardware connectivity primarily as a binary
relation: two qudits either can interact directly, or they cannot. In practice,
operations are not equally desirable across all physical locations or
interactions. Real devices exhibit heterogeneity in gate duration, error rate,
movement cost, and reliability. The Position Graph provides a natural interface
for representing this heterogeneity through edge and position labels, weights,
and capabilities. In this work, these quantities are treated as static during
compilation. A natural extension is to update them dynamically from calibration
data or time-varying device conditions, allowing routing decisions to adapt
without changing the underlying architecture abstraction.

The same abstraction can also represent operations beyond local movement or
nearest-neighbor gates. For example, teleportation-based interactions,
long-range entanglement links, or photonic interconnects could be modeled as
additional edge capabilities with distinct costs and constraints. Similarly,
frequently occurring multi-step procedures could be represented as composite
actions with precomputed costs. This would allow a routing algorithm to evaluate
higher-level transformations alongside primitive operations such as swaps,
moves, splits, and merges.

\section*{Acknowledgment}
This work was supported in part by NSF award \#2444042.

\bibliographystyle{unsrt}
\bibliography{bibliograph}

\end{document}